\documentclass[10pt, aps,pra, twocolumn,superscriptaddress]{revtex4-2}
\usepackage{amsmath,amssymb,amsfonts,bbm,graphicx,times,psfrag}
\usepackage[pdftex]{color}
\usepackage{graphicx}
\usepackage[colorlinks, linkcolor=blue, citecolor=blue, urlcolor=blue, breaklinks=true]{hyperref}
\usepackage{braket}
\usepackage{soul}
\usepackage{autobreak}
\usepackage{microtype}

\usepackage{amsthm}
\usepackage{csquotes}
\usepackage{amsthm}\newcommand{\be} {\begin{equation}}
\newcommand{\ee} {\end{equation}}

\usepackage{color}
\usepackage{mathrsfs}
\usepackage{amssymb}

\usepackage[skip=2pt]{caption}
\usepackage[utf8]{inputenc}
\DeclareUnicodeCharacter{2283}{\supset}
\usepackage{ragged2e}
\begin{document}
\title{Three-Photon and Hybrid Coherent-Fock Interference in a Two-Phase Six-Port Mach-Zehnder Interferometer}
\author{P. O. Amadi}
\email{amadiwati@gmail.com}
\affiliation{Division of Physical Science, Faculty of Science, Prince of Songkla University, Hat Yai, Songkhla 90110, Thailand}
\affiliation{Faculty of Electronic Engineering \& Technology, Universiti Malaysia Perlis, Arau, Perlis 02600, Malaysia}
\author{P. Pewkhom}
\affiliation{Division of Physical Science, Faculty of Science, Prince of Songkla University, Hat Yai, Songkhla 90110, Thailand}
\author{N. Sitpathom}
\affiliation{Division of Physical Science, Faculty of Science, Prince of Songkla University, Hat Yai, Songkhla 90110, Thailand}
\author{R. Endut}
\affiliation{Faculty of Intelligent Computing, Universiti Malaysia Perlis, Arau, Perlis 02600, Malaysia}
\affiliation{Centre of Excellence Advanced Communication Engineering (ACE), Universiti Malaysia Perlis,02600, Arau, Perlis, Malaysia}
\author{N. Ali}
\email{norshamsuri@unimap.edu.my}
\affiliation{Faculty of Electronic Engineering \& Technology, Universiti Malaysia Perlis, Arau, Perlis 02600, Malaysia}
\affiliation{Centre of Excellence Advanced Communication Engineering (ACE), Universiti Malaysia Perlis,02600, Arau, Perlis, Malaysia}
\author{S. A. Aljunid}
\affiliation{Faculty of Intelligent Computing, Universiti Malaysia Perlis, Arau, Perlis 02600, Malaysia}
\email{amadiwati@gmail.com}
\affiliation{Centre of Excellence Advanced Communication Engineering (ACE), Universiti Malaysia Perlis,02600, Arau, Perlis, Malaysia}
\author{S. Suryadi}
\email{suryadi008@binus.ac.id}
\affiliation{Computer Engineering Department, Faculty of Engineering, Bina Nusantara (BINUS) University, Kemanggisan, Palmerah, 11480, Jakarta, Indonesia}
\author{P. Kalasuwan}
\email{pruet.k@psu.ac.th}
\affiliation{Division of Physical Science, Faculty of Science, Prince of Songkla University, Hat Yai, Songkhla 90110, Thailand}

\begin{abstract}
We present a unified theoretical analysis of three-photon quantum interference in a Six-Port Mach-Zehnder Interferometer (6p-MZI) constructed from two cascaded tritters, with two independent phase modulators placed between the tritter arms. We analytically derive the transfer matrix of the 6p-MZI and show how they organize into three symmetry classes, governed by the discrete Fourier transform (DFT) structure of the tritter and the conjugate relations. Furthermore,  we analyze two input regimes: First, three indistinguishable single photons are injected into the tritter, and the output probability distributions $P_{[111]}$, $P_{[\{300\}]}$, and $P_{[\{210\}]}$ are derived as functions of the two relative phases $(\phi_1, \phi_2)$. At $\phi_2 = 0$, the single-phase limit is recovered, which confirms 100\% visibility of the even-distribution fringe. Second, a hybrid coherent-Fock input $|\alpha\rangle_1|\alpha\rangle_2|1\rangle_3$ is analyzed via the density matrix formalism. The average photon number at each output port exhibits amplitude-dependent phase shifts. Our results establish the 6p-MZI as a programmable platform for tripartite quantum state manipulation and coherent amplitude sensing. 
\end{abstract}
\maketitle

\section{Introduction}\label{sec:intro}

The principle of quantum interference lies at the heart of quantum mechanics and serves as a fundamental resource for emerging quantum 
technologies~\cite{zhong2020quantum,kok2007linear}. A most foundational manifestation of quantum interference is the Hong-Ou-Mandel (HOM) effect~\cite{hong1987measurement}: two indistinguishable photons at a balanced beam splitter interfere destructively, which suppresses all coincidence events. This purely quantum phenomenon, rooted in the bosonic symmetry of the two-photon wavefunction, has evolved from a tool for quantifying photon indistinguishability~\cite{faurby2024purifying,thomas2010measurement,yamamoto2003experimental} to becoming the cornerstone for photonic quantum computation~\cite{ladd2010quantum}, quantum teleportation~\cite{bouwmeester1997experimental}, quantum information processing~\cite{simon2016group}, quantum 
sensing~\cite{degen2017quantum}, and high-precision metrology~\cite{li2025stable,descamps2026role}.

The extension of these concepts to higher-dimensional Hilbert spaces through multiport devices are at the frontiers of intense theoretical and experimental interest~\cite{campos2000three,spagnolo2013three,kim2021implementation}. The symmetric three-port beam splitter, or tritter, provides the next-order generalization of the optical mixer. When compared with the standard beam splitter that is governed by SU(2) symmetry~\cite{campos1989quantum}, the tritter implements SU(3) unitary transformations that facilitate complex phenomena such as the three-photon bosonic coalescence~\cite{spagnolo2013three,mahrlein2015complete} and even-odd photon sorting~\cite{campos2000three}. Experimental realizations using integrated waveguides and single-mode optical fibers have confirmed that three-photon interference is sensitive to mixer phases and path reversals, and as such, offers an enhanced phase sensitivity when compared with two-path interferometers~\cite{lu2018electro, kim2021implementation,agne2017observation,pan2012multiphoton,vitanov2012synthesis,yan2025three, Weihs1996all}. Also, the connection to Boson sampling~\cite{aaronson2011computational,wang2018towards} 
further establishes multiport interference as a classically hard computational primitive that motivates exact analytical treatment of multi-photon statistics in these networks.

Therefore, motivated by the versatility of multiport mixers, the Six-Port Mach-Zehnder Interferometer (6p-MZI) has emerged as a programmable architecture for quantum state engineering~\cite{ripala2021hybrid, ke2022multiphoton}. This architecture is achieved by cascading two tritters with an intermediate phase 
modulator. Through heralded interference, the 6p-MZI implements programmable unitary operators that can manipulate multiphoton states~\cite{ke2022multiphoton, mane2025programmable}. Recent investigations have explored interference between hybrid combinations of Fock, coherent, and vacuum states~\cite{ripala2021hybrid,santiago2025photon}. In particular, the single-phase coherent-coherent-Fock (CCF) regime $|\alpha\rangle_1|\alpha\rangle_2|1\rangle_3$ has been analyzed in the 6p-MZI~\cite{suryadi7021493amplitude}. The analysis reveals amplitude-dependent phase shifts of the interference maxima that are absent in purely classical interference and arise from the weighted superposition of coherent-state and single-photon routing distributions. However, that analysis was restricted to a single phase modulator and did not address the three-photon Fock input or the two-dimensional phase space accessible with two independent modulators. The directionally-unbiased extension of this architecture was demonstrated experimentally in Ref.~\cite{kim2021implementation}, with two independent phase modulators $\phi_1$ and $\phi_2$. The experiment confirms the practical accessibility of the two-dimensional phase space. However, a unified analytical framework that exactly maps the symmetry classes of the 6p-MZI transfer matrix under two independent phase shifts, derives the complete two-dimensional coincidence probability distributions, and characterizes coherent amplitude-driven phase steering across the full output triplet is yet to be reported.

In this work, we address these gaps by extending the single-phase hybrid analysis of Ref.~\cite{suryadi7021493amplitude} to a full two-phase configuration. In particular, we derive the exact 6p-MZI transfer matrix and demonstrate that its elements organize into three symmetry triplets, governed by the DFT structure of the tritter and the conjugate relations $\omega^* = \omega^{-1}$, $\omega^{2*} = \omega$. We analyze two input regimes: First, in a pure quantum regime, three indistinguishable single photons $|1\rangle_1|1\rangle_2|1\rangle_3$ are injected, and the exact two-dimensional phase space of coincidence distributions is derived. We prove that the partial bunching channel is intrinsically insensitive to global phase shifts. Secondly, in the hybrid regime, we analyze the coherent-Fock input $|\alpha\rangle_1|\alpha\rangle_2|1\rangle_3$ via the density matrix formalism. We show that the dark-to-bright threshold is not fixed but depends on the phase trajectory through ($\phi_1, \phi_2$): it sits at $|\alpha|^2=2$ along the single-phase cut $\phi_2=0$, but drops to $|\alpha|^2=1/2$ along the symmetric phase $\phi_2=\phi_1$. We map this threshold surface across the full two-dimensional phase space and characterize the continuous amplitude-dependent phase steering at ports $1$ and $3$ along the same cuts. Our results establish 6p-MZI as programmable platform for tripartite quantum state manipulation and coherent amplitude sensing.

\section{The Six-Port Mach-Zehnder Interferometer}\label{sec:device}

\begin{figure}[ht!t!]
    \centering
    \includegraphics[width=0.95\linewidth]{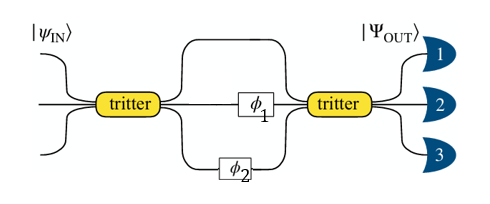}
    \caption{Schematic representation of the 6p-MZI. The device consists of two identical tritters (six-port beam splitters) that are separated by two independent phase modulators $\phi_1$ and $\phi_2$ 
acting on arms~2 and~3, with arm~1 fixed as the phase reference. Input states $|\psi_\mathrm{IN}\rangle=|1\rangle_1|1\rangle_2|1\rangle_3$ enter the first tritter, evolve into the intermediate state $|\Psi_1\rangle$, accumulate relative phases through $\hat{U}_\phi$. They interfere at the second tritter to produce the output state $|\Psi_\mathrm{OUT}\rangle$. The photon detection channels are represented by the output ports $1$, $2$, and $3$.}
    \label{fig:device}
\end{figure}

The optical system under investigation is the Six-Port Mach-Zehnder Interferometer (6p-MZI). As illustrated in Fig.~\ref{fig:device}, 6p-MZI comprises a first tritter as input coupler; two phase modulators acting on specific inter-tritter arms; and a second tritter as output coupler).

\subsection{The Ideal Tritter}\label{sec:tritter}

The core component of the 6p-MZI is the tritter, a six-port device. It comprises three inputs and three outputs and generalizes the function of a standard 50:50 beam splitter~\cite{spagnolo2013three,reck1994experimental, Weihs1996all}. We consider an ideal, lossless, and symmetric tritter whose unitary matrix, $\hat{U}_T$, in the DFT representation is

\begin{equation}
\hat{U}_T = \frac{1}{\sqrt{3}}
\begin{pmatrix}
1 & 1 & 1 \\
1 & \omega & \omega^2 \\
1 & \omega^2 & \omega
\end{pmatrix},
\quad
\omega = e^{i2\pi/3},\quad 
\omega^2 = e^{-i2\pi/3}.
\label{eq:tritter}
\end{equation}

The prefactor $1/\sqrt{3}$ ensures conservation of probability amplitude, and $\omega$ is the primitive cube root of unity, satisfying $1 + \omega + \omega^{-1} = 0$. This matrix structure ensures that a single photon entering any port exits with equal probability $1/3$ from any of the three output ports, with port-dependent phases that are essential for subsequent interference. The input-output relation for the creation operators $\hat{a}_j^\dagger$ (input) and $\hat{b}_i^\dagger$ (output) is

\begin{equation}
\hat{b}_i^\dagger = \sum_{j=1}^{3} (\hat{U}_T)_{ij}\hat{a}_j^\dagger,
\qquad [\hat{b}_i,\hat{b}_j^\dagger] = \delta_{ij},
\label{eq:forward}
\end{equation}

where the commutation relation confirms that the output particles remain bosons.
\subsection{The Composite 6p-MZI Transfer Matrix}\label{sec:transfer}

The total unitary transformation of the 6p-MZI is given by the ordered product of the individual optical transformations, which follows the physical propagation of light through the interferometer as

\begin{equation}
\hat{U}_{\rm MZI} = \hat{U}_T\hat{U}_\phi\hat{U}_T
= 
\begin{pmatrix}
u_{11} & u_{12} & u_{13} \\
u_{21} & u_{22} & u_{23} \\
u_{31} & u_{32} & u_{33}
\end{pmatrix}
\label{eq:MZI}
\end{equation}

Here, the element $u_{ij}$ is the probability amplitude for a photon that enters port~$j$ to exit through port~$i$. We compute $u_{ij}$ as

\begin{equation}
u_{ij} = \sum_{k=1}^{3}
(U_T)_{ik}\,e^{-i\phi_k^{\rm arm}}
(U_T)_{kj},
\label{eq:element}
\end{equation}
where $\phi_1^{\rm arm}=0$, $\phi_2^{\rm arm}=\phi_1$, $\phi_3^{\rm arm}=\phi_2$. The explicit forms of these coefficients, in Eq.~\ref{eq:MZI}, are obtained as follows:

\begin{align} \label{eq:classes}
u_{11} = u_{23} = u_{32} &= \frac{1}{3}
\bigl(1 + e^{-i\phi_1} + e^{-i\phi_2}\bigr)
\\ \nonumber
u_{12} = u_{21} = u_{33} &= \frac{1}{3}
\bigl(1 + e^{i(-\phi_1+\frac{2\pi}{3})} 
+ e^{-i(\phi_2+\frac{2\pi}{3})}\bigr)
\\ \nonumber
u_{13} = u_{31} = u_{22} &= \frac{1}{3}
\bigl(1 + e^{-i(\phi_1+\frac{2\pi}{3})} 
+ e^{i(-\phi_2+\frac{2\pi}{3})}\bigr)
\end{align}
Eq.~\ref{eq:classes} reveals that all the nine elements are organized into three symmetry classes. The triplet structure follows from two mechanisms: First, the DFT identity $\omega\cdot\omega^{-1}=1$ collapses specific intermediate path contributions before the phase modulator acts, and second, the conjugate relations $\omega^* = \omega^{-1}$ and $\omega^{2*} = \omega$ redistribute the diagonal elements across classes. The matrix is symmetric, $u_{ij} = u_{ji}$, which is consistent with reciprocity in a lossless optical network~\cite{potton2004reciprocity}. For single-phase limit, $\phi_2 = 0$, Eq.~\ref{eq:classes} reduces to

\begin{align}
u_{11} = u_{23} = u_{32} &= \frac{1}{3}
(2 + e^{-i\phi_1}),
\\ \nonumber
u_{12} = u_{21} = u_{33} &= \frac{1}{3}
\bigl(1 + e^{i(-\phi_1+\frac{2\pi}{3})} 
+ e^{-i\frac{2\pi}{3}}\bigr),
\\ \nonumber
u_{13} = u_{31} = u_{22} &= \frac{1}{3}
\bigl(1 + e^{-i(\phi_1+\frac{2\pi}{3})} 
+ e^{i\frac{2\pi}{3}}\bigr),
\end{align}
and at $\phi_1=0$ all the three classes reduce correctly to unity.

\section{Three-Photon Interference}\label{sec:fock}

In this section, we analyze the input state consisting of exactly 
one photon entering each port,

\begin{equation}\label{eq:inputs}
|\psi_{\rm IN}\rangle = |1\rangle_1|1\rangle_2
|1\rangle_3 = \hat{a}_1^\dagger\hat{a}_2^\dagger
\hat{a}_3^\dagger|000\rangle.
\end{equation}

Eq.~\ref{eq:inputs} is a three-photon generalization of the Hong-Ou-Mandel experiment~\cite{hong1987measurement}, and the indistinguishability of the three bosons drives interference phenomena. This is because they are governed by the permutation symmetry of the bosonic wavefunction.

\subsection{Intermediate State and Generalized HOM Effect}\label{sec:intermediate}

The first tritter transforms each input creation operator in the evolution direction as

\begin{equation}\label{eq:evolution}
\hat{a}_j^\dagger = \sum_{i=1}^3
(U_T^*)_{ij}\hat{b}_i^\dagger
\end{equation}

Note that Eq.~\ref{eq:evolution} makes use of $U_T^*$, the complex conjugate of the forward tritter matrix. This is because expressing input operators in terms of output operators requires 
inverting Eq.~\ref{eq:forward} via $\hat{U}_T^{-1} = \hat{U}_T^\dagger$. Therefore, Eq~\ref{eq:evolution} explicitly gives:

\begin{align}
\hat{a}_1^\dagger &\to \frac{1}{\sqrt{3}}
(\hat{b}_1^\dagger + \hat{b}_2^\dagger 
+ \hat{b}_3^\dagger),
\nonumber\\
\hat{a}_2^\dagger &\to \frac{1}{\sqrt{3}}
(\hat{b}_1^\dagger + \omega^{-1}\hat{b}_2^\dagger 
+ \omega\hat{b}_3^\dagger),
\nonumber\\
\hat{a}_3^\dagger &\to \frac{1}{\sqrt{3}}
(\hat{b}_1^\dagger + \omega\hat{b}_2^\dagger 
+ \omega^{-1}\hat{b}_3^\dagger).
\label{eq:transforms}
\end{align}

 Substituting Eq.~\ref{eq:transforms} into Eq.~\ref{eq:inputs} generates and organizes the terms into three output classes: the even distribution $|111\rangle$, 
the fully bunched NOON-like states 
$\{|300\rangle,|030\rangle,|003\rangle\}$, and 
the partially bunched states 
$\{|210\rangle,|201\rangle,|120\rangle,
|102\rangle,|021\rangle,|012\rangle\}$. The partially bunched class vanishes identically because its amplitude requires summing 
$1+\omega+\omega^{-1}=0$ over the three DFT phase paths. Again, this is a direct manifestation of the generalized Hong-Ou-Mandel effect~\cite{spagnolo2013three}. 

The surviving intermediate state after the first tritter is

\begin{equation}
|\Psi_1\rangle = -\frac{1}{\sqrt{3}}|111\rangle 
+ \frac{\sqrt{2}}{3}
\bigl(|300\rangle+|030\rangle+|003\rangle\bigr).
\label{eq:intermediate}
\end{equation}

The negative sign on $|111\rangle$ reflects destructive interference accumulated through the DFT phase paths and is a physical consequence of bosonic symmetry. 
 The $|210\rangle $-type coefficient is zero, which confirms HOM suppression. By the same argument, all six $|210 \rangle $-type states vanishes, since $1+\omega+\omega^{-1}=0$, always. 

\subsection{\label{sec:output}Output State and Probability Distributions}

The phase modulator $\hat{U}_\phi$ imprints a phase $e^{-i\phi_k^{\rm arm} n_k}$ on each Fock state component of $|\Psi_1\rangle$ according to its occupation of each arm, which gives

\begin{equation}
|\Psi_2\rangle = -\frac{e^{-i\Phi}}{\sqrt{3}}|111\rangle + \frac{\sqrt{2}}{3}
\bigl(|300\rangle + e^{-3i\phi_1}|030\rangle + e^{-3i\phi_2}|003\rangle\bigr),
\label{eq:phased}
\end{equation}
$\Phi = \phi_1+\phi_2$ is the total accumulated phase on the $|111\rangle$ component. In single-phase limit $\phi_2=0$, only $|030\rangle$ acquires a nontrivial phase factor $e^{-3i\phi_1}$, as expected physically, since only arm~$2$ is modulated.

The second tritter then acts on $|\Psi_2\rangle$ via the same evolution-direction transformation. Eq.~\ref{eq:evolution} is applied to $\hat{b}_k^\dagger$ operators (see Appendix~\ref{sec:appendix}). Computing the matrix for each component and tracking bosonic normalization factors $\sqrt{n!}$ throughout, the full output state is

\begin{align}
|\Psi_{\rm OUT}\rangle &= c_1|111\rangle_c + c_2\bigl(|300\rangle_c+|030\rangle_c
+|003\rangle_c\bigr)
\nonumber\\
&+ d_A\bigl(|210\rangle_c+|021\rangle_c+|102\rangle_c\bigr)
\nonumber\\
&+ d_B\bigl(|201\rangle_c+|120\rangle_c+|012\rangle_c\bigr),
\label{eq:output}
\end{align}

where the subscript $c$ labels the output-port creation operators $\hat{c}_k^\dagger$ after the second tritter. $c_1$ governs the even distribution; $c_2$ the fully bunched class; and $d_A$, $d_B$ the two distinct subclasses of partial bunching distinguished by the cyclic ordering of port indices. We define $S_3 = 1 + e^{-3i\phi_1} + e^{-3i\phi_2}$, the coefficients are

\begin{align}
c_1 &= \frac{e^{-i\Phi}}{3} + \frac{2S_3}{9},
\label{eq:c1}
\\
c_2 &= \frac{\sqrt{2}}{3\sqrt{3}}\left[-e^{-i\Phi} + \frac{S_3}{3}\right],
\label{eq:c2}
\\
d_A &= \frac{\sqrt{2}}{9}\bigl(1 + \omega^{-1}e^{-3i\phi_1} 
+ \omega e^{-3i\phi_2}\bigr),
\label{eq:dA}
\\
d_B &= d_A^*.
\label{eq:dB}
\end{align}

The conjugate relation $d_B = d_A^*$ follows directly from $\omega^* = \omega^{-1}$ and confirms the reciprocity of the network.

We calculate the detection probability for each output class, $P = |\text{coeff}|^2$, summed over all states in the class. Three physically distinct patterns emerge, expressed in terms of the pairwise sum

\begin{equation}
\mathcal{P} = \cos(3\phi_1) + \cos(3\phi_2) + \cos(3(\phi_1-\phi_2))
\label{eq:pairwise}
\end{equation}

and the cyclic sum

\begin{equation}
\mathcal{C} = \cos(\phi_1+\phi_2) + \cos(\phi_2-2\phi_1)
+ \cos(\phi_1-2\phi_2).
\label{eq:cyclic}
\end{equation}

The three probabilities are

\begin{align}
P_{[111]} &= \frac{7}{27} + \frac{8}{81}\mathcal{P} 
+ \frac{4}{27}\mathcal{C},
\label{eq:P111}
\\
P_{[\{300\}]} &= \frac{8}{27} + \frac{4}{81}\mathcal{P} - \frac{4}{27}\mathcal{C},
\label{eq:P300}
\\
P_{[\{210\}]} &= \frac{4}{27}(3 - \mathcal{P}).
\label{eq:P210}
\end{align}

The full normalization holds analytically:P$_{[111]} + P_{[\{300\}]} + P_{[\{210\}]} = 1$.

\subsection{Single-Phase Limit: \texorpdfstring{$\phi_2 = 0$}{φ₂ = 0}}\label{sec:single}

We set $\phi_2=0$ in Eqs.~(\ref{eq:P111})--(\ref{eq:P210}) gives 
$\mathcal{P}|_{\phi_2=0} = 2\cos(3\phi_1)+1$ and 
$\mathcal{C}|_{\phi_2=0} = 2\cos\phi_1 
+ \cos(2\phi_1)$, which yields

\begin{align}
P_{[111]} &= \frac{29}{81} + \frac{8}{27}\cos\phi_1 + \frac{4}{27}\cos(2\phi_1) + \frac{16}{81}\cos(3\phi_1),
\label{eq:P111_single}
\\
P_{[\{300\}]} &= \frac{28}{243} - \frac{8}{81}\cos\phi_1 
- \frac{4}{81}\cos(2\phi_1) + \frac{8}{243}\cos(3\phi_1),
\label{eq:P300_single}
\\
P_{[\{210\}]} &= \frac{8}{27}(1-\cos(3\phi_1)).
\label{eq:P210_single}
\end{align}

At $\phi_1=0$, $P_{[111]}=1$, $P_{[\{300\}]}=0$, $P_{[\{210\}]}=0$. The 100\% visibility of the $P_{[111]}$ fringe is a hallmark of perfect photon indistinguishability~\cite{spagnolo2013three}. The $P_{[\{210\}]}$ fringe carries a pure $\cos(3\phi_1)$ dependence, which represents a frequency tripling of the interference fringe relative to a standard two-port MZI, consistent with three-photon bosonic interference~\cite{campos2000three}.

\begin{figure}[ht!]
    \centering
    \includegraphics[width=0.95\linewidth]{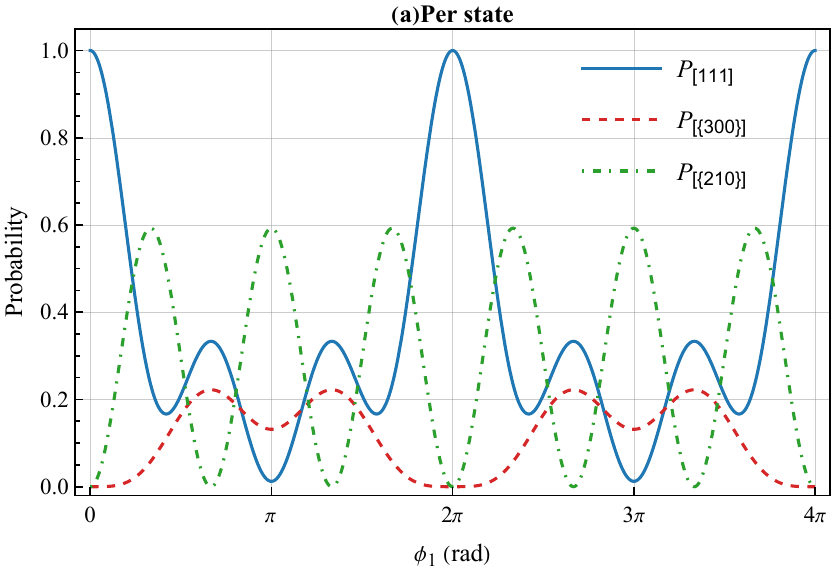}
    \includegraphics[width=0.95\linewidth]{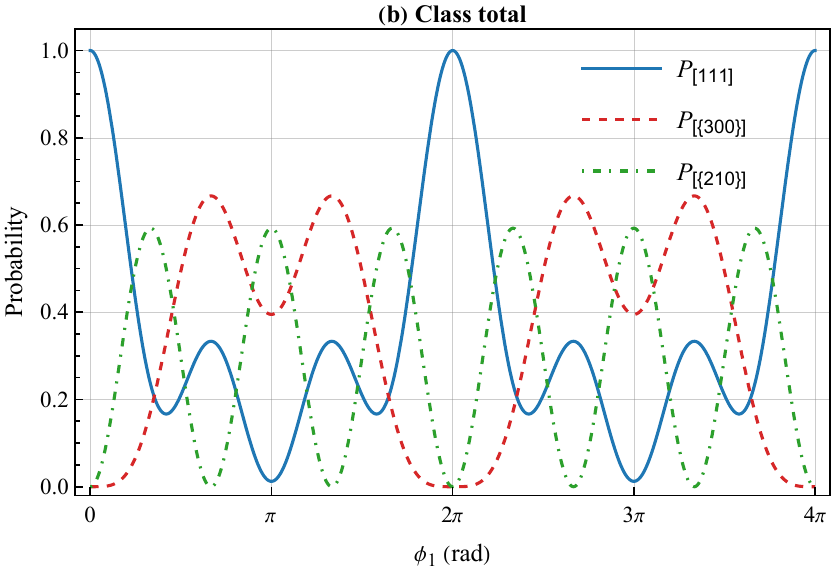}
    \caption{Detection probabilities for the three-photon input $|1\rangle_1|1\rangle_2|1\rangle_3$, in the single-phase limit $\phi_2=0$, plotted as a function of $\phi_1\in[0,4\pi]$. Panel~(a) shows the per-state probabilities and Panel~(b) shows the corresponding class total probabilities.$P_{[111]}$ (blue solid), $P_{[\{300\}]}$ (red dashed), and $P_{[\{210\}]}$ (green dash-dotted).}
    \label{fig:singlephase}
\end{figure}
Fig~\ref{fig:singlephase} shows $P_{[111]}$, $P_{[\{300\}]}$, and $P_{[\{210\}]}$ as functions of $\phi_1$ in the single-phase limit $\phi_2=0$. Panel~(a) shows the per-state probabilities and panel~(b) the class totals. The even-distribution fringe $P_{[111]}$ reaches unity at $\phi_1 = 2m\pi$ for integer $m$, reflecting the identity operation $\hat{U}_T^2=\hat{I}$ and confirming $100\%$ fringe visibility as a direct signature of perfect photon indistinguishability \cite{spagnolo2013three}. The partial bunching probability $P_{[\{210\}]}$ exhibits a pure $\cos(3\phi_1)$ dependence, representing a frequency tripling of the interference fringe relative to a standard two-port MZI; this is the hallmark of three-photon bosonic interference~\cite{campos2000three}. The full bunching probability $P_{[\{300\}]}$ is complementary to $P_{[\{111\}]}$, vanishing at $\phi_1=0$ and reaching its maximum where $P_{[\{111\}]}$ is suppressed, consistent with the opposite signs on the cyclic sum $\mathcal{C}$ in Eqs.~(\ref{eq:P111_single})--(\ref{eq:P300_single}).

\subsection{Two-Phase Analysis: Physical Consequences}\label{sec:twophase}

\begin{figure*}[ht!]
\centering
\includegraphics[width=2\columnwidth]{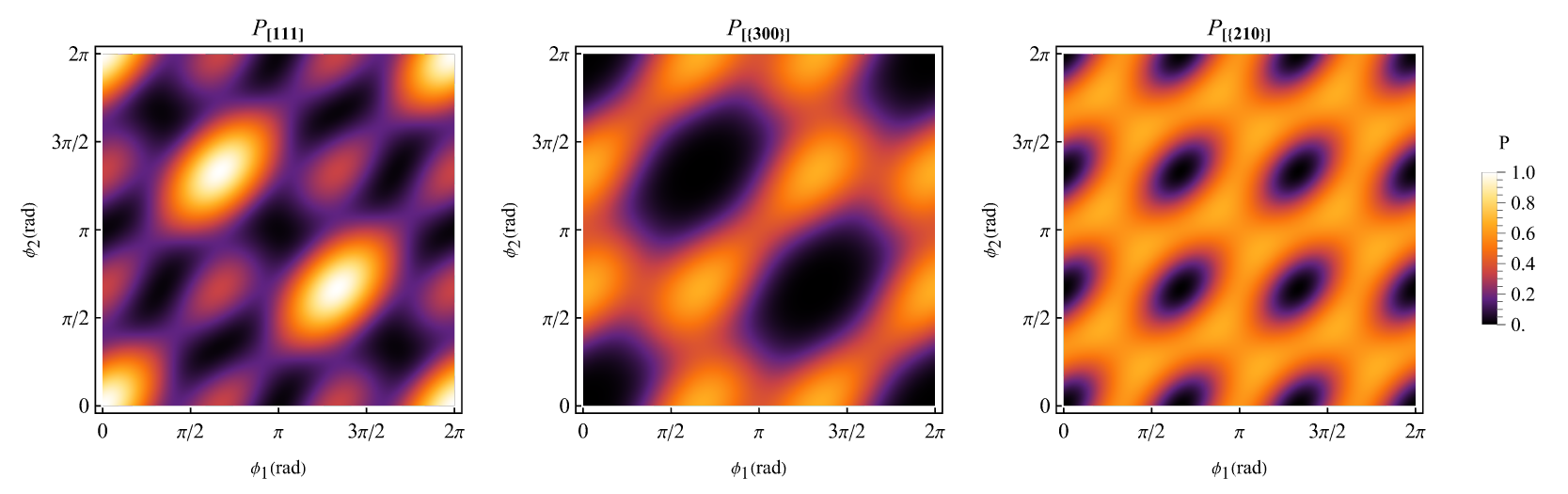}\\
\includegraphics[width=2 \columnwidth]{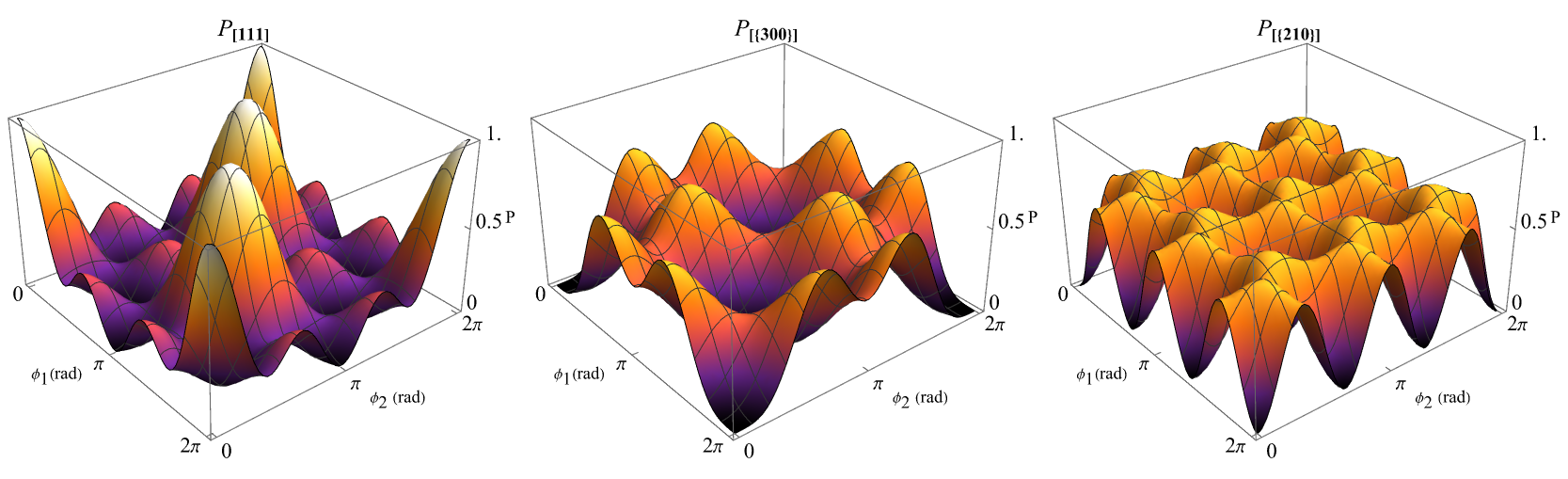}\\
\includegraphics[width=2 \columnwidth]{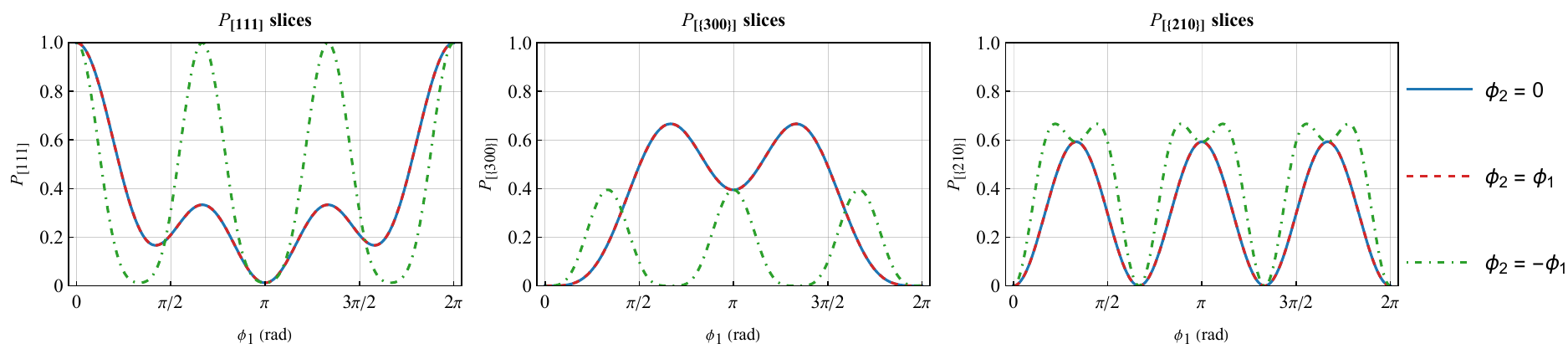} 
\caption{Two-dimensional phase space analysis of three-photon interference for two-phase configuration $(\phi_1,\phi_2)\in[0,2\pi]^2$. Top row: density maps of $P_{[111]}$, $P_{[\{300\}]}$, and $P_{[\{210\}]}$ over the full phase space; color encodes probability from 0 (dark) to 1 (bright). Middle row: corresponding 3D surface plots with the same color scale. Bottom row: one-dimensional slices along three physically motivated paths. The single-phase limit $\phi_2=0$ (blue solid), the symmetric configuration $\phi_2=\phi_1$ (red dashed), and the antisymmetric configuration $\phi_2=-\phi_1$ (green dash-dotted). In the antisymmetric slice, the cyclic sum $\mathcal{C}=0$ identically, decoupling $P_{[\{210\}]}$ completely from the other two channels.}
\label{fig:threephoton2d}
\end{figure*}

Figure~\ref{fig:threephoton2d} presents the complete two-dimensional phase-space analysis referenced throughout this subsection. Three physical features of the full two-dimensional phase space $(\phi_1,\phi_2)$ are revealed, which is not accessible in the single-phase limit.

\textit{First}, $P_{[\{210\}]}$ depends only on the pairwise sum $\mathcal{P}$. This contains only pairwise phase differences $\phi_1$, $\phi_2$, and $\phi_1-\phi_2$. The global phase $\Phi=\phi_1+\phi_2$ drops out entirely. This means the partial bunching channel is not sensitive to a uniform phase shift applied simultaneously to both arms. This symmetry is only invisible in the single-phase treatment, where pairwise and global contributions are locked to a single parameter.

\textit{Second}, $P_{[\{300\}]}$ and $P_{[111]}$ carry the same cyclic sum $\mathcal{C}$ with opposite signs. Therefore, $P_{[\{300\}]}$ and $P_{[111]}$ are strictly complementary: any phase configuration $(\phi_1,\phi_2)$ that drives coalescence suppresses even distribution and vice versa.

\textit{Third}, the two free parameters allow independent tuning of the pairwise and cyclic contributions. This cannot be achieved by $\phi_1$ alone. However, by setting $\phi_2=\phi_1$, the pairwise asymmetry vanishes and reduces $\mathcal{P}$ to $2\cos(3\phi_1)+1$ and $\mathcal{C}$ to 
$2\cos(\phi_1)+\cos(2\phi_1)$, which recovers a one-dimensional symmetric slice. We set $\phi_2=-\phi_1$ to create a fully antisymmetric configuration where $\mathcal{C}=0$ identically; this leaves $P_{[\{210\}]}$ as the only phase-dependent observable and decouples it completely from $P_{[111]}$ and $P_{[\{300\}]}$.

At $\phi_1=\phi_2=0$, all three probabilities are correctly reduced their identity-operation values. The $\cos(3(\phi_k-\phi_l))$ frequency tripling in $\mathcal{P}$, however, persists across the entire two-dimensional phase. This is a confirmation that the three-photon bosonic interference is the underlying mechanism regardless of how many relative phases are active. The two-phase configuration, therefore, adds the ability to engineer the interference landscape rather than simply sweep through it. This opens a direct route toward programmable quantum state preparation in tritter-based networks.

\section{Hybrid Coherent-Fock State Interference}
\label{sec:hybrid}

The coexistence of a coherent field and a Fock state inside the
interferometer creates a direct competition between the two distinct
sources of fluctuation: the coherent amplitudes which provide a classical phase backbone, while the injected single photon contributes quantum statistics~\cite{steindl2021artificial,barnett2022single}. 
Because the coherent-Fock configuration sits at the boundary between discrete- and continuous-variable quantum optics~\cite{huang2019engineering,sobhani2025nonclassicalities,xu2025entanglement,bartley2012multiphoton}, the 6p-MZI is uniquely positioned to probe how a single injected photon reshapes classical interference.

The input state is
\begin{equation}
|\psi_{\rm IN}\rangle = |\alpha\rangle_1|\alpha\rangle_2|1\rangle_3
= \hat D_1(\alpha)\hat D_2(\alpha)\hat a_3^\dagger|000\rangle,
\label{eq:inputccf}
\end{equation}
where $\hat D(\alpha)=\exp(\alpha\hat a^\dagger-\alpha^*\hat a)$ is the
displacement operator~\cite{walls2008quantum}. Because the 6p-MZI is a
linear network, the displacement operators transform covariantly under
it: $\hat U_{\rm MZI}\hat D(\alpha)\hat U_{\rm MZI}^\dagger =
\hat D(\alpha')$, with $\alpha'$ fixed by the transfer matrix of
Eq.~\eqref{eq:MZI}. by applying $\hat U_{\rm MZI}$ to Eq.~\eqref{eq:inputccf}, the
output state factorizes into a product of local displacements acting
on a single-photon superposition,
\begin{align}
|\Psi_{\rm OUT}\rangle = \hat D_1(\xi_1)\hat D_2(\xi_2)\hat D_3(\xi_3) \times \nonumber \\
\left(u_{13}\hat c_1^\dagger+u_{23}\hat c_2^\dagger+u_{33}\hat c_3^\dagger\right)|000\rangle.
\label{eq:outputs}
\end{align}
The three displacement operators carry the interferometrically
redistributed coherent fields. Also, the term in the bracket describes the single photon injected at port $3$, which coherently split across all three output modes with amplitudes $u_{13}$, $u_{23}$, $u_{33}$.

For $\alpha_1=\alpha_2=\alpha$ and $\alpha_3=0$, the output coherent
amplitude at port $k$ is the coherent sum of contributions
transmitted from input ports $1$ and $2$,
\begin{equation}
\xi_k = (u_{k1}+u_{k2})\,\alpha.
\label{eq:xik-def}
\end{equation}
By using the two-phase transfer-matrix elements of Sec.~\ref{sec:transfer}, and the with the
identities $1+\omega=-\omega^{-1}$, $1+\omega^{-1}=-\omega$, Eq.~\eqref{eq:xik-def} explicitly reads:
\begin{align}
\xi_1 &= \frac{\alpha}{3}\left(2-\omega^{-1}e^{-i\phi_1}-\omega e^{-i\phi_2}\right), \label{eq:xi1}\\
\xi_2 &= \frac{\alpha}{3}\left(2-e^{-i\phi_1}-e^{-i\phi_2}\right), \label{eq:xi2}\\
\xi_3 &= \frac{\alpha}{3}\left(2-\omega e^{-i\phi_1}-\omega^{-1}e^{-i\phi_2}\right). \label{eq:xi3}
\end{align}
Two structural features of Eq.~\eqref{eq:xi1}-\eqref{eq:xi3} are worth noting. First, $\xi_3$ is obtained from $\xi_1$ by exchanging $\omega\leftrightarrow\omega^{-1}$. This reflects the conjugate symmetry $\omega^*=\omega^{-1}$ that relates ports $1$ and $3$ in the transfer matrix. Second, at $\phi_1=\phi_2=\phi$, the central-port amplitude is reduced to $\xi_2 = (\alpha/3)(2-2e^{-i\phi})$, which vanishes at $\phi=0$ and reaches its maximum modulus at $\phi=\pi$.

\subsection{Average photon number}\label{sec:hybrid-A}

The mean photon number at port $k$ follows directly from
Eq.~\eqref{eq:output}:
\begin{equation}
\langle n_k\rangle = |\xi_k|^2 + |u_{k3}|^2,
\label{eq:nk-general}
\end{equation}
where the first term is the coherent contribution and the second term is the contribution arising from single-photon injected at port $3$. From a direct evaluation of $\langle\Psi_{\rm OUT}|\hat n_k|\Psi_{\rm OUT}\rangle$, confirms that no cross term appears between the two sources, at any $(\phi_1,\phi_2)$ - the coherent and single-photon parts of the mean photon number are strictly additive. From Eq.~\eqref{eq:nk-general}, we evaluate the two-phase transfer matrix, at all three ports, which reads:
\begin{widetext}
\begin{align}
\langle n_1\rangle &= \frac{1}{9}\Big[6|\alpha|^2+3
+(4|\alpha|^2-2)\cos\!\big(\phi_1-\tfrac{\pi}{3}\big)
+(4|\alpha|^2-2)\cos\!\big(\phi_2+\tfrac{\pi}{3}\big)
-(2|\alpha|^2+2)\cos\!\big(\phi_1-\phi_2+\tfrac{\pi}{3}\big)\Big], \label{eq:n1}\\
\langle n_2\rangle &= \frac{1}{9}\Big[6|\alpha|^2+3
+(2-4|\alpha|^2)\cos\phi_1
+(2-4|\alpha|^2)\cos\phi_2
+(2|\alpha|^2+2)\cos(\phi_1-\phi_2)\Big], \label{eq:n2}\\
\langle n_3\rangle &= \frac{1}{9}\Big[6|\alpha|^2+3
+(4|\alpha|^2-2)\cos\!\big(\phi_2-\tfrac{\pi}{3}\big)
+(4|\alpha|^2-2)\cos\!\big(\phi_1+\tfrac{\pi}{3}\big)
-(2|\alpha|^2+2)\cos\!\big(\phi_1-\phi_2-\tfrac{\pi}{3}\big)\Big], \label{eq:n3}
\end{align}
\end{widetext}
which satisfies the photon-number conservation, $\langle n_1\rangle+\langle n_2\rangle+\langle n_3\rangle = 2|\alpha|^2+1$, for every $(\phi_1,\phi_2)$. This is a direct consequence of the DFT identity $\cos\theta+\cos(\theta\pm2\pi/3)=0$. Apparently, all single-phase results of Ref.~\cite{suryadi7021493amplitude} are recovered at $\phi_2=0$. Also, we compare Eqs.~\eqref{eq:n1}--\eqref{eq:n3}, which also gives the exact port-exchange identity
\begin{equation}
\langle n_3\rangle(\phi_1,\phi_2) = \langle n_1\rangle(\phi_2,\phi_1),
\label{eq:port-exchange}
\end{equation}
which is a direct consequence of the conjugate relation $\omega^*=\omega^{-1}$ that is linking ports $1$ and $3$. Equation~\eqref{eq:port-exchange} implies that the two outer ports carry identical two-dimensional interference information that is related by a fixed coordinate exchange; not by two independent functions. Therefore, every result derived below for port $1$ applies to port $3$ under $\phi_1\leftrightarrow\phi_2$. 
Furthermore, the fingerprint of the two-phase configuration is observed in the cross terms $\cos(\phi_1-\phi_2\pm\pi/3)$ and $\cos(\phi_1-\phi_2)$, which is entirely absent from any single-phase analysis. If we set $\phi_2=0$, it collapses each cross term into a constant that reshapes, rather than supplements the single-modulator.

\subsection{The dark-to-bright threshold}\label{sec:hybrid-B}

Equations~\eqref{eq:n1}--\eqref{eq:n3} share a single structural
feature: every cosine coefficient is built from the same factor
$(2-4|\alpha|^2)$, up to a port-dependent sign. At port 2 the factor
multiplies $\cos\phi_1$ and $\cos\phi_2$ directly; at ports 1 and 3 it
appears with the opposite sign, multiplying the corresponding
phase-shifted cosines. This common factor changes sign at the
critical amplitude
\begin{equation}
|\alpha|^2_{\rm th} = \frac{1}{2},
\label{eq:threshold}
\end{equation}
independent of which port is observed. This is different from the single-phase case, where only port 2 exhibits a dark-to-bright transition~\cite{suryadi7021493amplitude}. Also, the critical amplitude occurs at the four-times-larger amplitude $|\alpha|^2=2$, set by the SU(3) symmetry class, by placing $u_{23}$ in the same triplet as $u_{11}$ and $u_{32}$. The factor-of-four reduction arises because $\phi_1$ and $\phi_2$ contribute independently to the coherent amplitude that reaches every output port and effectively doubling the coherent power that competes against the fixed single-photon routing term $|u_{k3}|^2$.

To make the threshold explicit, we evaluate Eq.~\eqref{eq:n2} along the two phase trajectories. Along the symmetric cut $\phi_2=\phi_1\equiv\phi$,
\begin{equation}
\langle n_2\rangle\Big|_{\phi_2=\phi_1} = \frac{1}{9}\Big[(8|\alpha|^2+5)+(4-8|\alpha|^2)\cos\phi\Big],
\label{eq:n2-sym}
\end{equation}
which has twice the modulation depth of the corresponding
single-phase expression. The oscillating coefficient $(4-8|\alpha|^2)$ changes sign at $|\alpha|^2=1/2$: below threshold it is positive and $\langle n_2\rangle$ dips below unity at $\phi=0,2\pi$; above threshold the sign reverses and $\langle n_2\rangle$ instead peaks at
those same phases. At threshold, $\langle n_2\rangle=1$ identically, and independent of $\phi$, which mark a reversal between dark and bright operating points. Along the antisymmetric cut $\phi_2=-\phi_1$, the differential phase $\phi_1-\phi_2=2\phi\equiv -\phi$, introduces an additional harmonic,
\begin{align}
\langle n_2\rangle\Big|_{\phi_2=-\phi_1} = \frac{1}{9}\Big[(6|\alpha|^2+3)+2(2-4|\alpha|^2)\cos\phi+ \nonumber \\
(2|\alpha|^2+2)\cos(2\phi)\Big].
\label{eq:n2-antisym}
\end{align}
Here, only the $\cos\phi$ term inverts sign at threshold, while the $\cos(2\phi)$ term persists unchanged across it. This produces a richer and two-frequency interference profile than the symmetric cut.

\begin{figure}[!ht]
\centering
\includegraphics[width=\columnwidth]{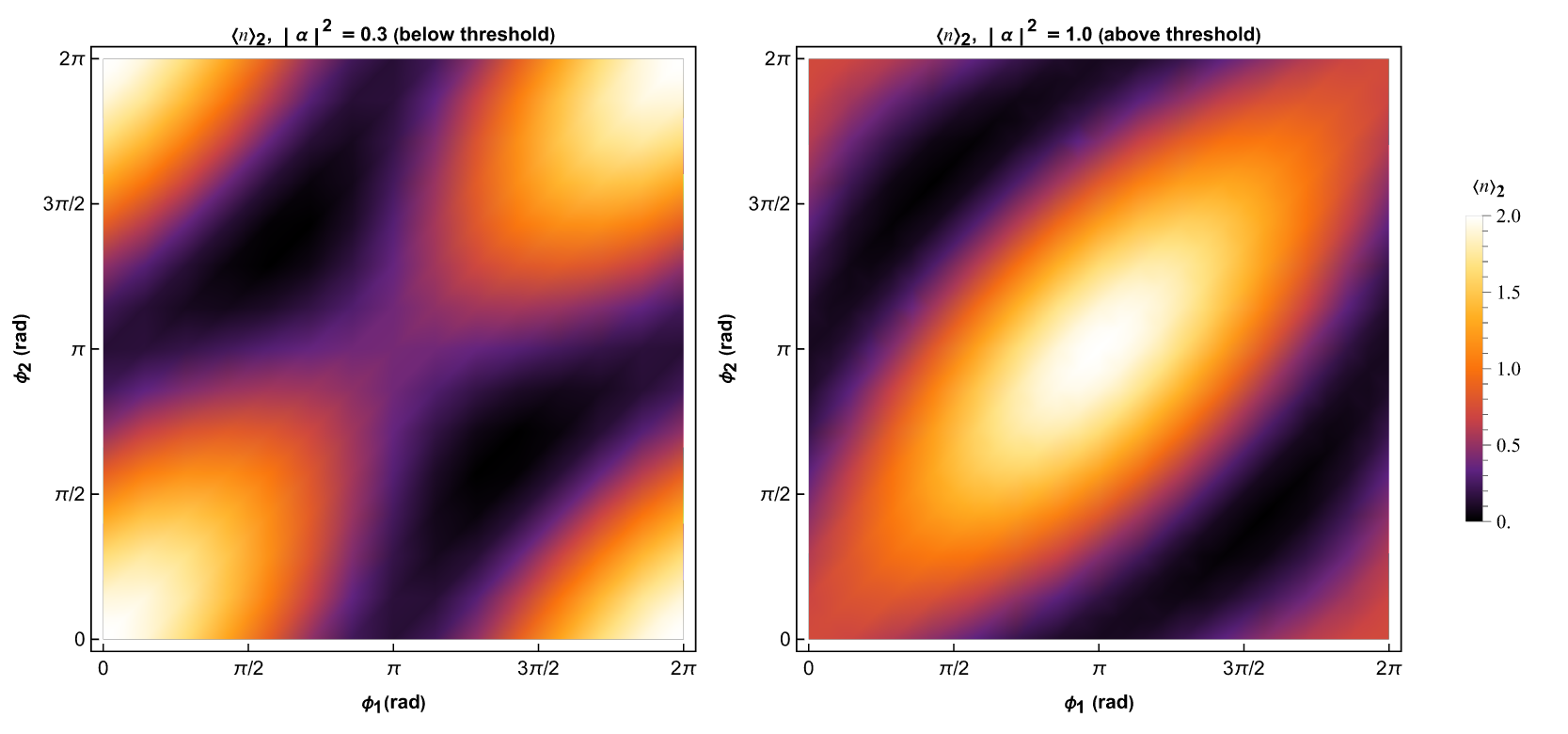}
\caption{Density maps of $\langle n_2\rangle$ over the full
two-dimensional phase space $(\phi_1,\phi_2)\in[0,2\pi]^2$, for
$|\alpha|^2=0.3$ (below threshold) and $|\alpha|^2=1.0$ (above
threshold).} 
\label{fig:n2_density}
\end{figure}

Figure~\eqref{fig:n2_density} shows the resulting two-dimensional landscape
$\langle n_2\rangle(\phi_1,\phi_2)$ below and above threshold. Below threshold, the density map displays diagonal stripes that run parallel to $\phi_1=\phi_2$. The signature of the $\cos(\phi_1-\phi_2)$ term dominate the destructive interference along that diagonal. Above threshold, the pattern reorganizes into a single elliptical bright lobe, centered around $(\pi,\pi)$. This qualitative reorganization, rather than a simple rescaling, confirms that the sign flip of $(2-4|\alpha|^2)$ reshapes the full two-dimensional
\begin{figure}[!ht]
\centering
\includegraphics[width=\columnwidth]{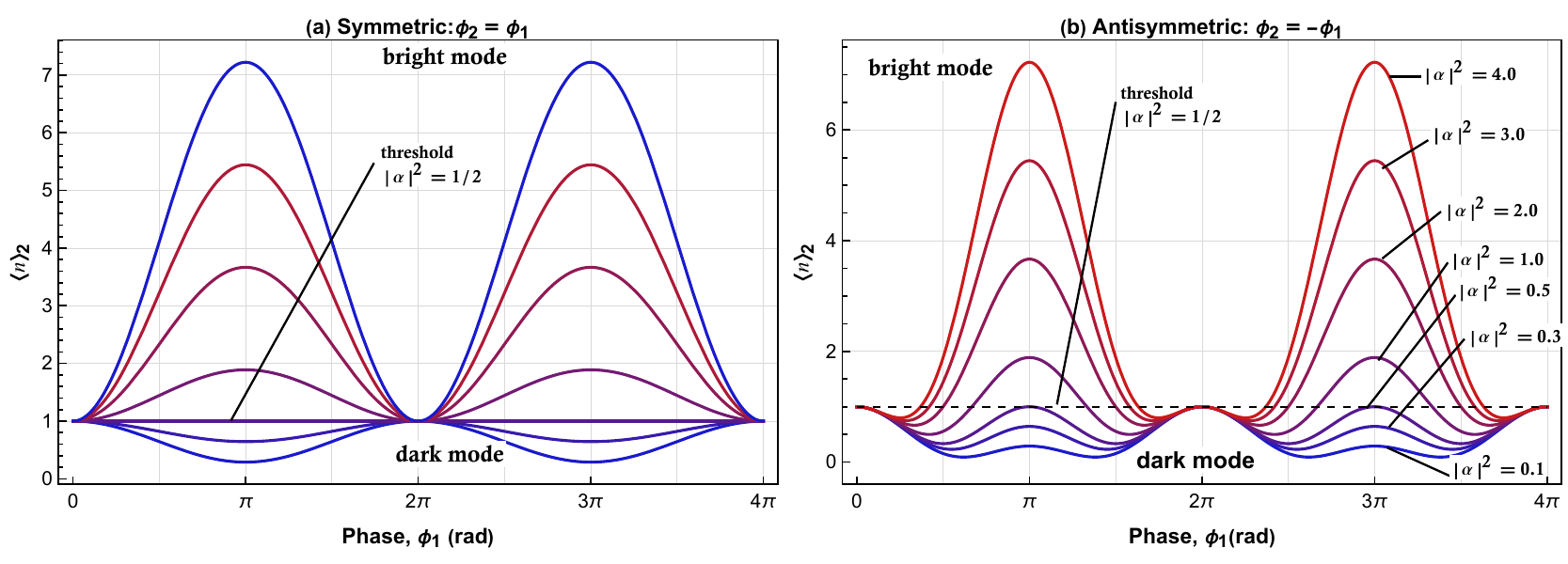}
\caption{$\langle n_2\rangle$ versus $\phi_1$ along the (a) symmetric phase $\phi_2=\phi_1$ and (b) antisymmetric phase $\phi_2=-\phi_1$, for seven values of $|\alpha|^2$ from 0 to 4.0. The solid line (right) and the dashed line (left) along $\langle n_2 \rangle$, at $|\alpha|^2=1/2$, marks the threshold where the phase dependence inverts.}
\label{fig:n2_cuts}
\end{figure}

Figure~\eqref{fig:n2_cuts} follows $\langle n_2\rangle$ along symmetric and antisymmetric cuts for seven amplitudes, spanning $|\alpha|^2=0$ to $4.0$, with the analytic threshold of Eq.~\eqref{eq:threshold}. The symmetric cut directly confirms Eq.~\eqref{eq:n2-sym}, while the antisymmetric cut shows the same qualitative inversion, however, with a modified curve shape from the surviving $\cos(2\phi)$ term of Eq.~\eqref{eq:n2-antisym}.
 
Conceptually, this dark-to-bright transition is related to the intensity-based discrimination between a macroscopic reference pulse and a sub-single-photon signal in the B92 quantum-key-distribution  protocol~\cite{bennett1992quantum}; the 6p-MZI generalizes that binary balance to three spatial output modes with a continuously tunable operating point. This mechanism is qualitatively different from B92, where the discrimination threshold is fixed by a 50:50 beam-splitter condition. In our architecture, it is set entirely by the DFT symmetry of the SU(3) transfer matrix and by the number of active phase modulators, which makes $|\alpha|^2_{\rm th}$ a tunable device parameter rather than a fixed design constant. 

\subsection{Amplitude-dependent phase steering at the outer ports}
\label{sec:hybrid-C}

The three-term structure of Eqs.~\eqref{eq:n1} and~\eqref{eq:n3}
produces phase steering at the outer ports that is considerably
richer than in the single-phase case, and that depends on which cut through $(\phi_1,\phi_2)$ is taken. The interference-maximum condition at port 1, from Eq.~\eqref{eq:n1}, reads
\begin{equation}
(2-4|\alpha|^2)\sin\!\big(\phi_1-\tfrac{2\pi}{3}\big)
+(2|\alpha|^2+2)\sin\!\big(\phi_1-\phi_2+\tfrac{2\pi}{3}\big) = 0,
\label{eq:n1-max-general}
\end{equation}
which couples $\phi_1$ and $\phi_2$ and must, in general, be solved jointly. 
\begin{figure*}[ht!]
\centering
\includegraphics[width=0.99 \linewidth]{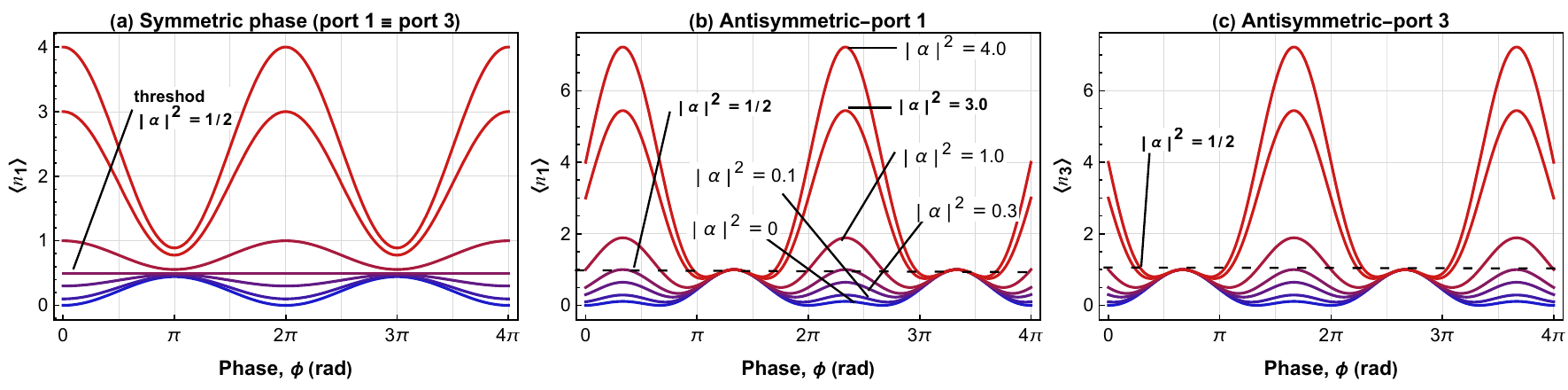}
\caption{$\langle n_{1,3}\rangle$ swept over one phase at fixed values of the other, for $|\alpha|^2\in\{0,0.1,0.3,0.5,1.0,3.0,4.0\}$. (a) Port 1 swept over $\phi_1$ at fixed $\phi_2$ which coincides exactly with port 3 swept over $\phi_2$ at fixed $\phi_1$. (b) Port 1 swept over $\phi_1$ at fixed $\phi_1$ which coincides exactly with port 3 swept over $\phi_1$ at fixed $\phi_2$.}
\label{fig:independent_sweep}
\end{figure*}

Figure~\eqref{fig:independent_sweep}(a) confirms, by direct evaluation of Eqs.~\eqref{eq:n1} and~\eqref{eq:n3}, that $\langle n_1\rangle =\langle n_3\rangle$ exactly along the symmetric $\phi_2=\phi_1$. This is consistent with Eq.~\eqref{eq:port-exchange} which is algebraically equivalent to driving a single modulator on arm 1 alone. Furthermore, along the antisymmetric cut $\phi_2=-\phi_1$ in Figs.~\eqref{fig:independent_sweep}(b,c), this degeneracy is not present ($\langle n_1\rangle\neq\langle n_3\rangle$). This is because the sign reversal $\phi_2\to-\phi_2$ does not follow the conjugate relation that links the two ports. Both diagonal cuts remain confined to a single effective phase, however, and neither can, by construction, reveal how the interference landscape depends on $\phi_1$ and $\phi_2$ independently.

\begin{figure}[ht!]
\centering
\includegraphics[width=\columnwidth]{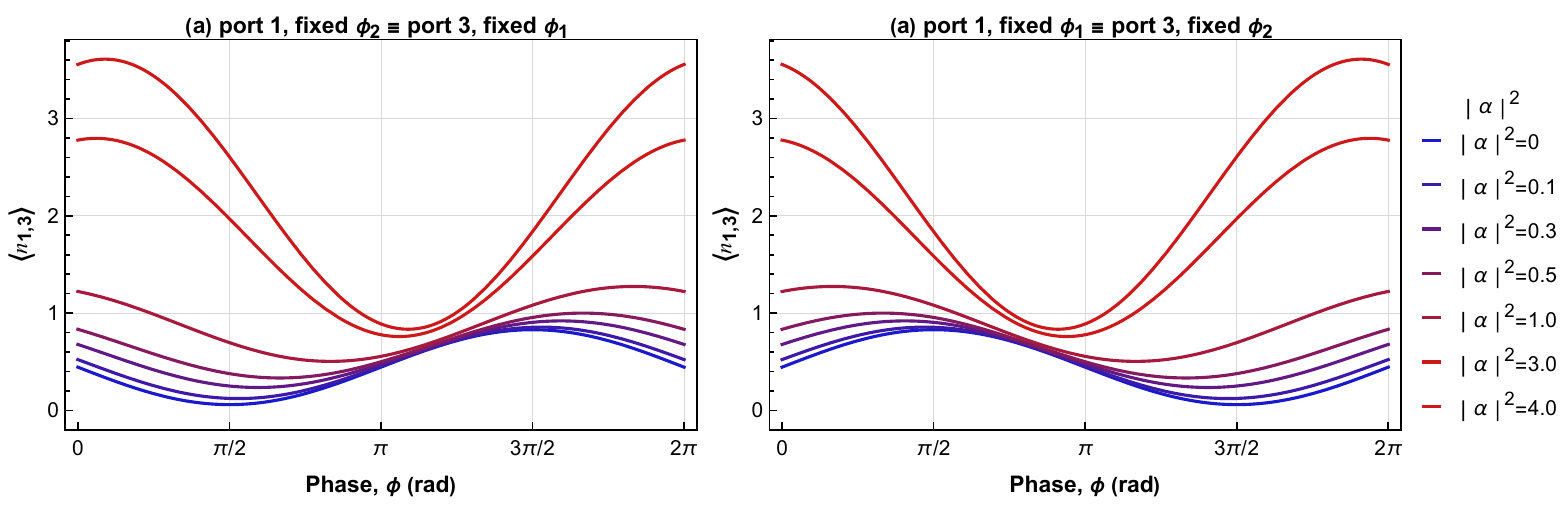}
\caption{$\langle n_{1,3}\rangle$ swept over one phase at fixed values of the other, for $|\alpha|^2\in\{0,0.1,0.3,0.5,1.0,3.0,4.0\}$. (a) Port 1 swept over $\phi_1$ at fixed $\phi_2$ coincides exactly with port 3 that is swept over $\phi_2$ at fixed $\phi_1$. (b) Port 1 swept over $\phi_1$ at fixed $\phi_1$ coincides exactly with port 3 that is swept over $\phi_1$ at fixed $\phi_2$.}
\label{fig:independent_sweeps}
\end{figure}

Figure~\ref{fig:independent_sweeps} shows that once the two phases are
swept independently. In this case, we observe that the maximum $\langle n_1\rangle$ and $\langle n_3\rangle$ drifts continuously with
$|\alpha|^2$, rather than flipping discretely between two fixed
values as it does on the symmetric and antisymmetric cuts. The reason
 can be seen in Eq.~\eqref{eq:n1}: $\phi_1$ and $\phi_2$ enter
through different phase offsets, $\cos(\phi_1-\pi/3)$ and
$\cos(\phi_2+\pi/3)$, therefore, fixing one of the phase and sweeping the other, samples a single harmonic whose phase offset differs from panel to panel of Fig.~\ref{fig:independent_sweeps} and not the common offset that collapses both diagonal cuts to a shared threshold. Because the
maximum condition, Eq.~\eqref{eq:n1-max-general}, then mixes two
harmonics with different phase offsets; and no single crossing amplitude analogous to $|\alpha|^2_{\rm th}=1/2$ exists along
these sweeps. Of the four combinations, that can be constructed from $\{\text{port }1,\text{port }3\}\times\{\text{sweep }\phi_1,\text{sweep }\phi_2\}$, only two are independent. However, no information is lost by this reduction to two curve families and the remaining two combinations are recovered from Fig.~\ref{fig:independent_sweeps} by relabeling which phase is held as fixed.

\begin{figure}[!ht]
\centering
\includegraphics[width=\columnwidth]{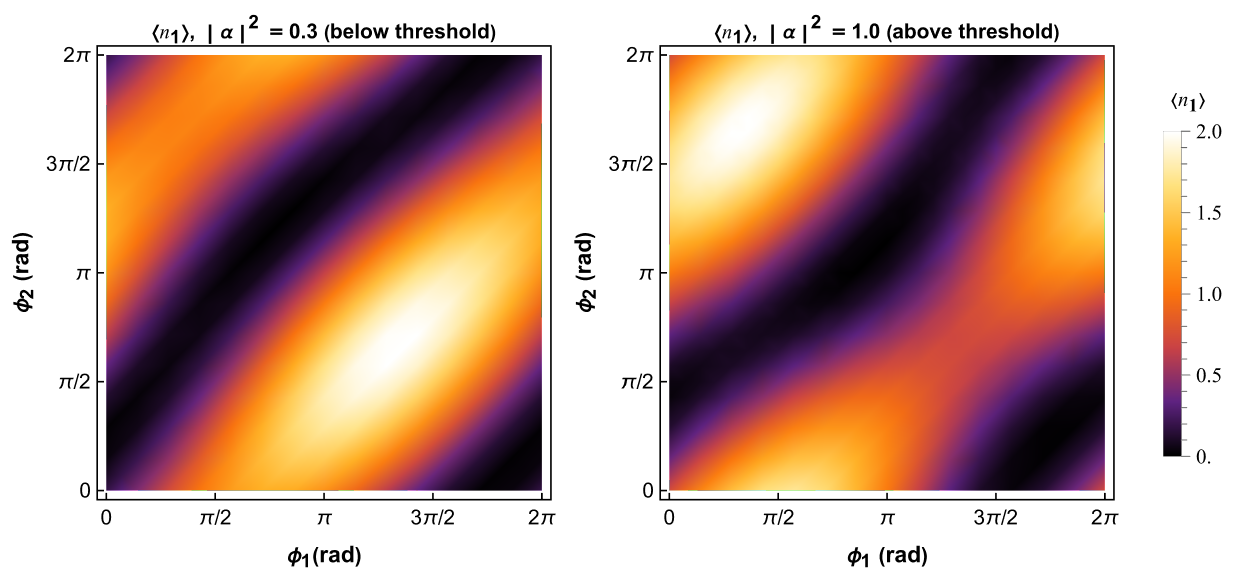}
\caption{$\langle n_1\rangle$ over the full independent phase space $(\phi_1,\phi_2)\in[0,2\pi]^2$ that is below ($|\alpha|^2=0.3$) and above ($|\alpha|^2=1.0$) the threshold. The bright ridge runs parallel to the diagonal $\phi_1=\phi_2$ below threshold and reorganizes into a single lobe above it. This is consistent with the sign reversal of the shared coefficient $(2-4|\alpha|^2)$ identified in Sec.~\ref{sec:hybrid-B}.}
\label{fig:n1_landscape}
\end{figure}

Figure~\ref{fig:n1_landscape} shows $\langle n_1\rangle$ over the complete $(\phi_1,\phi_2)$ plane at the same two representative amplitudes as in Fig.~\ref{fig:n2_density}. Below threshold ($|\alpha|^2=0.3$), the bright ridge runs parallel to the diagonal $\phi_1=\phi_2$. The diagonal is  the trajectory along which the general two-phase form, Eq.~\eqref{eq:n1}, collapses into single-harmonic symmetric-cut result. Below threshold, the coefficient $(4|\alpha|^2-2)$ is negative and the maximum sits at $\phi=\pi$. Therefore, the ridge in the 2D map traces out that maximum as $\phi_1=\phi_2$ is varied together. Above threshold, the same coefficient changes sign, the maximum on the diagonal moves to $\phi=0\ (\mathrm{mod}\ 2\pi)$, and the extended ridge collapses into the single compact lobe visible in the right panel. 
The corresponding map for $\langle n_3\rangle$ is the mirror reflection of Fig.~\ref{fig:n1_landscape} across the diagonal $\phi_1=\phi_2$. 

We now examine the two-phase trajectories: Along the symmetric slice $\phi_2=\phi_1\equiv\phi$, the port-1 photon number collapses to a single harmonic,
\begin{equation}
\langle n_1\rangle\Big|_{\phi_2=\phi_1} = \frac{1}{9}\Big[5|\alpha|^2+2+(4|\alpha|^2-2)\cos\phi\Big],
\label{eq:n1-sym}
\end{equation}
with extrema at $\phi=0$ and $\phi=\pi$, which is determined by the sign of $(4|\alpha|^2-2)$. The interference maximum sits at $\phi=\pi$ for $|\alpha|^2<1/2$ and at $\phi=0$ for $|\alpha|^2>1/2$, and flips at $|\alpha|^2=1/2$ threshold. At threshold, the phase-dependent term vanishes and $\langle n_1\rangle=\langle n_3\rangle=1/2$ identically and independent of $\phi$. The symmetric trajectory therefore reproduces the same threshold-induced $\pi$-phase inversion at the outer ports as at the central port, which confirming that Eq.~\eqref{eq:threshold} governs all three ports simultaneously rather than being a port-2 peculiarity. Only the photon number at threshold differs between the center port ($\langle n_2\rangle=1$) and outer ports ($\langle n_1\rangle,\langle n_3\rangle=1/2$), since the total $\langle n_1\rangle+\langle n_1\rangle+\langle n_3\rangle=2|\alpha|^2+1=2$ at $|\alpha|^2=1/2$ must be
shared unequally between them.

Along the antisymmetric slice $\phi_2=-\phi_1\equiv-\phi$, and by contrast, the cross term does not collapse to a constant. This is because $\phi_1-\phi_2=2\phi$. Instead, it contributes a genuine second harmonic,
\begin{align}
\langle n_1\rangle\Big|_{\phi_2=-\phi_1} = \frac{1}{9}\Big[6|\alpha|^2+3
+2(4|\alpha|^2-2)\cos\!\big(\phi-\tfrac{\pi}{3}\big) \nonumber \\
-(2|\alpha|^2+2)\cos\!\big(2\phi+\tfrac{\pi}{3}\big)\Big],
\label{eq:n1-antisym}
\end{align}
with extrema fixed by
\begin{equation}
(2-4|\alpha|^2)\sin\!\big(\phi-\tfrac{\pi}{3}\big)
+(2|\alpha|^2+2)\sin\!\big(2\phi+\tfrac{\pi}{3}\big) = 0.
\label{eq:n1-antisym-max}
\end{equation}
Unlike the symmetric cut, Eq.~\eqref{eq:n1-antisym-max} mixes a
fundamental and a second-harmonic term with competing $|\alpha|^2$ dependence. Therefore, there is a continuous shift of the stationary-point location with $|\alpha|^2$, rather than flipping between two fixed values. At $|\alpha|^2=1/2$ the first-harmonic term vanishes and the extrema are set entirely by the surviving second harmonic. Away from threshold, the both terms contribute, and the maximum steers continuously between
these two regimes. This is the genuinely two-phase result: the
symmetric cut reproduces a discrete threshold that is shared with port 2, while the antisymmetric cut is where the outer ports display steering behavior, with no single-phase or symmetric-cut analogue.

\section{Conclusion}\label{sec:conclusion}

This work presents an analysis of three-photon quantum interference in the 6p-MZI. We investigate two physically distinct input regimes under a two-phase modulator configuration that spans the full programmable phase space of the 6p-MZI device. For a pure quantum input $|1\rangle_1|1\rangle_2|1\rangle_3$, we analytically derive and show that the transfer matrix can be organized into three symmetry triplets: $\{u_{11},u_{23},u_{32}\}$, 
$\{u_{12},u_{21},u_{33}\}$, and $\{u_{13},u_{31},u_{22}\}$. 
Here, we derived the output probability distributions $P_{[111]}$, $P_{[\{300\}]}$, and $P_{[\{210\}]}$ as exact functions of the pairwise sum $\mathcal{P}$ and cyclic sum $\mathcal{C}$ over the two-dimensional phase space $(\phi_1,\phi_2)$. Second, the hybrid coherent-Fock input $|\alpha\rangle_1|\alpha\rangle_2|1\rangle_3$ were analyzed, via the displacement operator formalism. The average photon number at each port separates into a coherent contribution $|\xi_k|^2$, and a single-photon contribution $|u_{k3}|^2$ with no cross terms. The two-phase treatment shows the dark-to-bright threshold at port 2 is not a fixed property of the device but a function of the phase trajectory, falling from $|\alpha|^2=2$ at $\phi_2=0$ in Ref.~\cite{suryadi7021493amplitude} to $|\alpha|^2=1/2$ along $\phi_2=\phi_1$. This tunability is itself a metrological resource, since it lets an experimenter choose the operating amplitude at which phase sensitivity vanishes or peaks

The results of our analysis have direct experimental relevance. The two-phase modulator configuration analyzed here is architecturally identical to the directionally-unbiased 3$\times$3 fiber multiport demonstrated by Kim et al.~\cite{kim2021implementation}. Their investigation achieved transfer matrix reconstruction with fidelity $F=0.971\pm0.005$ using entangled photon pairs from SPDC at $1556$~nm. The three-photon coincidence distributions derived here, $P_{[111]}$, $P_{[\{300\}]}$, and $P_{[\{210\}]}$, can be directly measured with superconducting nanowire single-photon detectors (SNSPDs) of the type used in~\cite{kim2021implementation}. Also, the $\cos(3\phi_1)$ frequency tripling in $P_{[\{210\}]}$ provides a phase sensitivity enhancement of factor~3 over a standard two-port MZI. The hybrid regime results are accessible with coherent states from a continuous-wave laser attenuated to the few-photon level, which can be combined with a heralded single photon from SPDC; the $|\alpha|^2=1/2$ dark-to-bright threshold corresponds to a mean photon number readily achieved with milliwatt-level pump powers in current fiber-optic implementations. Future work will examine the effect of photon distinguishability, which includes temporal and spectral mismatch, on the interference visibility of the distributions derived in this work; the generalization to $N$-photon inputs in SU($N$) networks; and the application of the two-phase programmable interference landscape to multi-parameter quantum sensing beyond the standard quantum limit~\cite{Humphreys2013quantum, Giovannetti2011advances}.

\section*{Acknowledgements}
This research has received funding support from the NSRF via the Research and Innovation Acceleration Agency for Competitiveness and Area 4/5 Development (RCAD) (Program Management Unit for Frontier Brainpower and Future Industries) [grant number B39G690076]

\appendix
\section{Mode Transformations Under the Second Tritter}\label{sec:appendix}

The second tritter applies $\hat{U}_T$ to $|\Psi_2\rangle$. The intermediate mode operators $\hat{b}_k^\dagger$ lives between the two tritters and the output mode operators $\hat{c}_k^\dagger$ lives after the second tritter. The forward direction of Eq.~(\ref{eq:forward}) defines how output operators are built from input operators:

\begin{equation}
\hat{c}_i^\dagger = \sum_{k=1}^3 (U_T)_{ik}\hat{b}_k^\dagger.
\label{eq:app_forward}
\end{equation}

Inverting via $\hat{U}_T^{-1}=\hat{U}_T^\dagger$:

\begin{equation}
\hat{b}_k^\dagger = \sum_{i=1}^3
(U_T^*)_{ik}\hat{c}_i^\dagger,
\label{eq:app_inverse}
\end{equation}

where the conjugate tritter matrix is

\begin{equation}
U_T^* = \frac{1}{\sqrt{3}}
\begin{pmatrix}
1 & 1 & 1 \\
1 & \omega^{-1} & \omega \\
1 & \omega & \omega^{-1}
\end{pmatrix}.
\label{eq:app_conjugate}
\end{equation}

Reading off column by column:

\begin{align}
\hat{b}_1^\dagger &\to \frac{1}{\sqrt{3}}(\hat{c}_1^\dagger+\hat{c}_2^\dagger
+\hat{c}_3^\dagger),
\nonumber\\
\hat{b}_2^\dagger &\to \frac{1}{\sqrt{3}}(\hat{c}_1^\dagger+\omega^{-1}\hat{c}_2^\dagger
+\omega\hat{c}_3^\dagger),
\nonumber\\
\hat{b}_3^\dagger &\to \frac{1}{\sqrt{3}}
(\hat{c}_1^\dagger+\omega\hat{c}_2^\dagger
+\omega^{-1}\hat{c}_3^\dagger).
\label{eq:app_transforms}
\end{align}

The structure is identical to the first tritter transformations Eq.~(\ref{eq:transforms}). This reflects to the fact that both tritters are of the same physical device. Therefore, the evolution-direction transformation always reads columns of $U_T^*$, regardless of which tritter is being applied. The two applications differ only in which operators are being transformed: 
$\hat{a}_k^\dagger\to\hat{b}_k^\dagger$ at the 
first tritter and 
$\hat{b}_k^\dagger\to\hat{c}_k^\dagger$ at the 
second.

The bosonic commutation relations are preserved: 
for $i=j=1$,

\begin{equation}
[\hat{c}_1,\hat{c}_1^\dagger] = \frac{1}{3}
(1+1+1) = 1,
\label{eq:app_comm1}
\end{equation}

and for $i\neq j$, say $i=1$, $j=2$:

\begin{equation}
[\hat{c}_1,\hat{c}_2^\dagger] = \frac{1}{3}
(1+\omega^{-1}+\omega) = \frac{1}{3}(0) = 0,
\label{eq:app_comm2}
\end{equation}

using $1+\omega+\omega^{-1}=0$, which confirms that the output operators remain bosonic.


%

\end{document}